\documentclass[prd,twocolumn]{revtex4}
\usepackage{graphicx}
\usepackage{epsfig}
\usepackage{color}
\usepackage{mathrsfs}
\usepackage{amsmath}

\newcommand{\be}{\begin{equation}}
\newcommand{\ee}{\end{equation}}
\newcommand{\bd}{\begin{equation*}}
\newcommand{\ed}{\end{equation*}}
\newcommand{\bea}{\begin{eqnarray}}
\newcommand{\eea}{\end{eqnarray}}

\newcommand{\gapp}{\mathrel{\raise.3ex\hbox{$>$}\mkern-14mu
              \lower0.6ex\hbox{$\sim$}}}
\newcommand{\lapp}{\mathrel{\raise.3ex\hbox{$<$}\mkern-14mu
              \lower0.6ex\hbox{$\sim$}}}

\begin{document}

\title{Pre-Hawking Radiation from a Collapsing Shell}
\author{Eric Greenwood}
\author{Dmitry Podolsky}
\author{Glenn Starkman}
\affiliation{CERCA, Department of Physics, Case Western Reserve University, Cleveland, OH 44106-7079}
\begin{abstract}
We investigate the effect of induced massive radiation given off during the time of collapse of a massive spherically symmetric domain wall in the context of the functional Schr\"odinger formalism. Here we find that the introduction of mass suppresses the occupation number in the infrared regime of the induced radiation during the collapse. The suppression factor is found to be given by $e^{-\beta m}$, which is in agreement with the expected Planckian distribution of induced radiation. Thus a massive collapsing domain wall will radiate mostly (if not exclusively) massless scalar fields, making it difficult for the domain wall to shed any global quantum numbers and evaporate before the horizon is formed.
\end{abstract}

\maketitle

\section{Introduction}

Black holes embody one of the greatest frontiers in theoretical physics -- the intersection of quantum theory with general relativity. So far, attempts to confront this intersection and produce a single theory incorporating both quantum mechanics and general relativity, have brought along with them several interesting properties (such as Hawking radiation, see \cite{Hawking}), solidified existing arguments (such as black hole entropy, see \cite{Bekenstein}), and  introduced unanswered paradoxes associated with the formation and evaporation of a black hole. These include the information loss  due to the thermal nature of Hawking evaporation (see \cite{Hawking2,Preskill,Mathur,Giddings}).

One radical resolution of the information loss paradox was recently proposed in \cite{VachStojKrauss}, where the collapse of a thin spherically symmetric domain wall (shell) was studied by means of the functional Schr\"odinger formalism (see also \cite{GreenStoj,Green}). Solving the Wheeler-de Witt equation for the collapsing shell, the authors of \cite{VachStojKrauss} argued that (a) such collapse takes an infinite amount of time from the point of view of a distant observer, even when quantum fluctuations of the shell are taken into account (but ignoring any back-reaction due to the emission of radiation), and (b) the shell emits pre-Hawking massless radiation prior to the formation of a trapped surface. This pre-Hawking radiation (PHR) could, they found, cause the shell to evaporate completely  in  time,  $t_{\rm evap}\sim{}R_S \left(\frac{R_S}{l_P}\right)^2$ (measured from the moment when the radiation begins to effectively carry away  the mass of the shell). If  (a) and (b) are both true, then  no actual black hole can be formed in such a collapse, and, although the quantum state of  PHR from the collapsing shell is entangled with the quantum state of the shell, information about the initial quantum state of the shell is continuously released during the process of evaporation. They found that the spectrum of  PHR is approximately thermal~\footnote{The effective temperature of induced radiation emitted by a shell of mass $M$ is of the same order of magnitude as the Hawking temperature $T\sim{}\frac{M_P^2}{M}$ of a black hole of the same mass.} only at sufficiently large frequencies $\omega R_s\gg 1$, where $R_s$ is the Schwarzschild radius of the collapsing shell. The infrared part $\omega{}R_s\ll 1$ of the spectrum, on the other hand, remains time-dependent and does not exhibit the Rayleigh-Jeans singular behavior characteristic of the Hawking radiation in finite time. The infrared part of the spectrum also depends on the initial quantum state of the shell and preserves all the information encoded in it.

In the present paper we argue, to the contrary, that PHR will not prevent the formation of a trapped surface in the collapse of a spherical shell with mass $M\gg{}M_P$. First, we take into account quantum fluctuations of the Schwarzschild radius $R_S$ (``quantum trembling'' of the event horizon, see \cite{Frolov}) of the shell, present the correction terms to the functional Schr\"odinger equation describing the quantum collapse of the spherical shell and show that such corrections make the collapse time finite and in fact rather short: $t_{\rm coll}\sim{}R_S$, counted from the moment when pre-Hawking evaporation starts being effective. Second, we show that the actual evaporation time $t_{\rm evap}$ is significantly longer than the time scale $t_{\rm evap}\sim{}R_S \left(\frac{R_S}{l_P}\right)^2$ for pre-Hawking evaporation through emission of massless quanta, once one takes into account that the main contribution to the mass of the shell is given by constituent protons and neutrons. The latter implies that a typical collapsing shell has very large baryon/lepton quantum numbers, and in order for the shell to evaporate completely prior to collapse, these quantum numbers must either be erased by physical processes, or radiated away to infinity. Since global quantum numbers, such as $B$ and $L$, are only carried by quanta of \emph{massive} quantum fields, such as leptons and baryons, the relevant time scale to be compared to the total collapse time is the time of evaporation due to emission of massive  PHR.

We extend the analysis of \cite{VachStojKrauss} to the case of  PHR of massive particles and find the corresponding evaporation time. For masses of known leptons and baryons (including likely values of the neutrino masses) and astrophysically interesting masses of the collapsing shell $\frac{M}{M_P}\gg\frac{M_P}{m}\gg 1$ the total evaporation time is found to be significantly longer than the collapse time $t_{\rm coll}$.  We note how  non-perturbative Standard Model baryon-number violation softens, but does not eliminate, this conclusion.

This paper is organized as follows. In the Sec. \ref{sec:collapsetime} we construct the functional Schr\"odinger equation describing collapse of a quantum spherically symmetric shell and find corrections to this equation induced by fluctuations of the metric. We then compare classical and quantum collapse times for the shell. In Sec. \ref{sec:massive} we analyze massive  PHR emitted by the collapsing shell by means of function Schr\"odinger formalism, derive its spectrum and estimate its backreaction on the process of collapse. In Sec. \ref{sec:BLviolation} we discuss the effect of baryon and lepton number violation on the process of collapse. Finally, the Sec. \ref{sec:conclusion} is devoted to the discussion and conclusions.

\section{Estimates of the Collapse Timescale}
\label{sec:collapsetime}

\subsection{Collapse Time in Classical Physics}
\label{sec:classical}

Following \cite{VachStojKrauss} let us consider a perfectly spherical domain wall representing a spherical shell of collapsing matter, so that the position of the shell is completely described by its radius, $R(t)$. Our first goal is to check the well-known statement that from the point of view of a distant observer, the collapse takes infinite amount of time. Since we neglect any quantum effects, such as vacuum fluctuations, the Schwarzschild radius of the shell, $R_s=2GM$, remains unchanged during the course to collapse, see \cite{Ipser}.

Due to the spherical symmetry of the configuration and Birkhoff's Theorem, the metric can be naturally separated into two different regions. The metric of the space-time exterior to the collapsing shell is (see \cite{Ipser}) Schwarzschild:
\begin{equation}
ds^2= -\left(1-\frac{R_s}{r}\right) dt^2 + \frac{dr^2}{1-R_s/r}+
      r^2 d\Omega^2 \ , \ \ r > R(t)
\label{metricexterior}
\end{equation}
where as usual
\begin{equation}
d\Omega^2  = d\theta^2  + \sin^2\theta d\phi^2 \, .
\end{equation}
In the interior of the shell, the metric is flat
\begin{equation}
ds^2= -dT^2 +  dr^2 + r^2 d\theta^2  + r^2 \sin^2\theta d\phi^2  \ ,
\ \ r < R(t)
\label{metricinterior}
\end{equation}
The interior time coordinate, $T$, is related to the asymptotic observer time coordinate, $t$, via the proper time of an observer moving with the shell, $\tau$. The relations are
\begin{equation}
\frac{dT}{d\tau} =
      \sqrt{1 + \left (\frac{dR}{d\tau} \right )^2}
\label{bigTandtau}
\end{equation}
and
\begin{equation}
\frac{dt}{d\tau} = \frac{1}{B} \sqrt{B+\left ( \frac{dR}{d\tau} \right )^2}
\label{littletandtau}
\end{equation}
where
\begin{equation}
B(t) \equiv 1 - \frac{R_s}{R(t)}.
\label{BofR}
\end{equation}

By taking the ratio of (\ref{bigTandtau}) and (\ref{littletandtau}), the relationship between the interior time $T$ and the asymptotic time $t$ is found to be
\be
  \frac{dT}{dt}=\frac{\sqrt{1+R_{\tau}^2}B}{\sqrt{B+R_{\tau}^2}}=\sqrt{B-\frac{(1-B)}{B}\dot{R}^2}\,,
  \label{bigTandlittlet}
\ee
where $R_{\tau}=dR/d\tau$ and $\dot{R}=dR/dt$.

Integrating the equations of motion for the shell gives an  expression for the total mass of the shell \cite{Ipser}, which is the integral of motion,
\begin{equation}
M = \frac{1}{2} [ \sqrt{1+R_\tau^2} + \sqrt{B+ R_\tau^2} ] 4\pi \sigma R^2 \,.
\label{ISmass}
\end{equation}
Here $\sigma$ is the surface tension of the shell. This expression for $M$ is implicit since $R_s =2GM$ occurs in $B$. Solving for $M$ explicitly in terms of $R_\tau$ gives
\begin{equation}
M = 4\pi \sigma R^2 [ \sqrt{1+R_\tau^2} - 2\pi G\sigma R] .
\label{MRtau}
\end{equation}

When $|R-R_s|/R_s \ll 1$, i.e. when the shell is close to its Schwarzschild radius,
the velocity of the shell as observed by an asymptotic observer, is given by
\be
  \dot{R}\approx-B\,.
  \label{Rdot}
\ee
Thus, to lowest order in $(R_0-R_S)$,
\be
  R(t)\approx R_S+(R_0-R_S)e^{-t/R_S}
  \label{R}
\ee
where $R_0=R(t=0)$ is the initial position of the shell \cite{VachStojKrauss}.

As we see, classical collapse of a spherically symmetric shell indeed takes infinite amount of time from the point of view of an asymptotic observer, although the distance between the shell and its Schwarzschild radius quickly becomes exponentially small. In what follows, we will count time starting from the moment where the behavior of the radius of the collapsing shell is approximately given by (\ref{R}).

\subsection{Collapse Time in Quantum Physics}
\label{sec:quantum}

We shall now consider how the classical estimate for the collapse time is modified once quantum effects (quantum fluctuations of the collapsing shell and of  the metric) are taken into account.

As shown in \cite{VachStojKrauss}, the collapse time will remain infinite in the presence of quantum fluctuations in the position of the collapsing shell. Let us recall how exactly one comes to this conclusion.  First, one assumes that the quantum dynamics of the system (the collapsing shell, the background space-time and the distant observer) are completely described by the Wheeler-de Witt equation
\begin{equation}
H\Psi=0
\label{eq:WdWTanmay}
\end{equation}
with the Hamiltonian
\begin{equation}
H=H_{{\rm sys}}+H_{{\rm obs}},
\end{equation}
where $H_{{\rm obs}}$ is the Hamiltonian of the distant observer. Since the proper distance between the collapsing shell and the observer is very large, the quantum state of the observer is not entangled with the state of the shell and the total wave function $\Psi$ is separable:
\begin{equation}
\Psi=\sum_{k}c_{k}\Psi_{k}^{{\rm sys}}\Psi_{{\rm obs}}^{k}.
\label{Psi_eq}
\end{equation}
One also expects that the state of observer $\Psi^{{\rm obs}}$ satisfies the usual Schr\"odinger equation
\begin{equation}
i\frac{\partial\Psi^{{\rm obs}}}{\partial t}=H_{{\rm obs}}\Psi^{{\rm obs}},
\end{equation}
where $t$ is the observer's time coinciding with the usual Schwarzschild time. Therefore, the quantum state of the collapsing shell should also satisfy Schr\"odinger's equation
\begin{equation}
i\frac{\partial\Psi^{{\rm sys}}}{\partial t}=H_{{\rm sys}}\Psi^{{\rm sys}}.
\label{eq:FuncSchEq}
\end{equation}
The  Hamiltonian of the shell is \cite{VachStojKrauss}
\begin{equation}
H_{{\rm sys}}^{2}=B\Pi\cdot B\Pi+B(4\pi\mu R^{2})^{2},
\label{eq:HamiltonianQuadratic}
\end{equation}
where $\Pi=-i\frac{\partial}{\partial R}$ is the canonical momentum, $B$ is given by (\ref{BofR}) and $\mu\equiv\sigma(1-2\pi G\sigma R_{S})$ is a constant. In the near horizon limit $R\to R_{S}$, only the ``kinetic'' term survives in (\ref{eq:HamiltonianQuadratic}), and the ``squared'' functional Schr\"odinger equation (\ref{eq:FuncSchEq}) becomes especially easy to solve using tortoise coordinates $t$,$u=R+R_{S}\log|R/R_{S}-1|$:
\begin{equation}
\left(\frac{\partial^{2}}{\partial t^{2}}-\frac{\partial^{2}}{\partial u^{2}}\right)\Psi^{{\rm sys}}\approx0.
\label{eq:KleinGordon}
\end{equation}
Its solution describes a wave packet propagating with the speed of light, with its width remaining fixed in the tortoise coordinates. Note however that the horizon is located at $u=-\infty$, so that it takes infinite time for the wave packet to reach it.

The considerations above do not, however, take all possible physical effects into account. For example, in deriving (\ref{eq:FuncSchEq}), we assume {\it a priori} that neither the horizon scale $R_{S}$, nor the mass of the shell $M=\frac{R_{S}}{2G}$ fluctuates. However, both should fluctuate, at least when we try to measure their value with precision finer than the Planck scale, simply because the metric of space-time itself strongly fluctuates at scales of the order of Planck length $l_{P}$. Naively, it seems that these tiny Planckian fluctuations cannot be relevant for the physics discussed because the curvature of space-time at the scale $R\approx R_{S}$ is significantly smaller than $M_{P}^{2}$. Recall however that the size of the shell $R$ approaches the Schwarzschild radius $R_{S}$ exponentially quickly (see (\ref{R})), so that the difference $R-R_{S}\sim R_{0}e^{-t/R_{S}}$ soon becomes comparable to the Planckian length $l_{P}$. Therefore, at time scales $t\gg R_{s}\log(R_{0}/l_{P})$ fluctuations of the metric must be taken into account.

How can one then account for these fluctuations at the level of the Wheeler-de Witt equation in (\ref{eq:WdWTanmay})? By separating the metric fluctuations in the full Wheeler-de Witt equation from the fluctuations of the matter degrees of freedom associated with the collapsing shell and distant observer, one can write
\begin{align}
H\Psi=\left(-16\pi G\cdot G_{ab,cd}\frac{\delta^{2}}{\delta g_{ab}\delta g_{cd}}-\frac{1}{16\pi G}\sqrt{g}R^{(3)}+\right.\nonumber\\
\left.+H_{{\rm sys}}+H_{{\rm obs}}\right)\Psi=0,
\label{eq:WdWfull}
\end{align}
where $g_{ab}$ is $3$-metric of the space-time, $R^{(3)}$ is $3$-curvature and $G_{ab,cd}=g^{-1/2}(g_{ac}g_{bd}+g_{ad}g_{bc}-g_{ab}g_{cd})$ is the local de Witt supermetric. The functional Schrodinger equation \cite{Banks}, as well as corrections to it \cite{KieferSingh}, can be constructed from (\ref{eq:WdWfull}) by looking for solutions of the form
\begin{equation}
\Psi=e^{iS}
\end{equation}
with the ``effective action'' $S$ expanded in powers of $G$:
\begin{equation}
S=G^{-1}S_{0}+S_{1}+GS_{2}+\ldots.
\label{eq:Gexp}
\end{equation}
The leading term $S_{0}$ describes the classical background space-time and the classical collapsing shell, and equations of motion derived by variating ``the action'' $S_0$ are Einstein-Jacobi equations for the background. The term $S_{1}$ takes into account (a) the vacuum fluctuations of geometry and (b) fluctuations of the shell and their backreaction on the background geometry. Variation of $S_{1}$ gives the functional Schr\"{o}dinger equation discussed above. Finally, the term $S_{2}$ and higher order terms describe the backreaction of the vacuum fluctuations of geometry on the background as well as interaction between fluctuations of the shell and fluctuations of geometry. Equations of motion found by variation of $S_{2}$ and higher order terms represent corrections to the functional Schr\"odinger equation.

The original Wheeler-de Witt equation (\ref{eq:WdWfull}) is notoriously hard to solve for many reasons, one of them being that it requires regularization in the same way as does the Schr\"odinger equation describing behavior of the wave functional of relativistic quantum fields --- behavior of a quantum field theory can be understood much more effectively by using second quantization and Feynman diagrammatics than first quantization and the Schr\"odinger equation. In fact, it is natural to expect that the very expansion in (\ref{eq:Gexp}) breaks down at the vicinity of event horizon $R\to R_{S}$, when the behavior of the radius of the shell is approximately given by the expression in (\ref{R}) and where, as we shall see, the evaporation process seems to be the most effective.

Fortunately, the relevant physics of the perturbation theory in powers of $G$ can be understood using uncertainty relations. In particular, in order to estimate the collapse time one can notice that the horizon is expected to ``tremble'' at Planckian scales \cite{trembling}, since the fluctuations of the metric $g$ can be estimated as
\begin{equation}
\frac{\delta g}{g}\sim\frac{l_{P}^{2}}{L^{2}},
\end{equation}
where $L$ is the length scale characterizing the curvature of the space-time \footnote{Note that the trembling effect cannot be gauged away, since tensor modes also contribute to it.} \cite{trembling}. Correspondingly, the characteristic size of fluctuation of the Schwarzschild radius can be estimated as
\begin{equation}
\delta R_{S}\sim\frac{l_{P}^{2}}{R_{S}}.
\label{eq:QuantumTrembling}
\end{equation}
Recall again that at $t\gg R_{S}$ the radius of the shell $R$ is exponentially close to the Schwarzschild radius $R_{s}$, and due to the uncertainty in (\ref{eq:QuantumTrembling}) in determining the value of $R_{S}$, the observer at infinity will be unable to determine whether the actual event horizon is formed or not after the time
\begin{equation}
t\sim R_{s}\log\left(\frac{R_{s}R_{0}}{l_{P}^{2}}\right)
\label{eq:CollapseTime}
\end{equation}
which can therefore be understood as the collapse time of the shell.

Another way to reproduce the estimate of (\ref{eq:CollapseTime}) is to use the fact that a distant observer can only determine the position of the collapsing shell by sending quanta towards it, having them scattered off the shell and measuring the corresponding cross sections (scattering and/or gravitational capture). Apparently, the maximal energy of the quantum, which the observer can send to the shell, is of the order of $M_{P}$. In the vicinity of the shell the blueshifted maximal energy of the quantum will therefore be bounded by
\begin{equation}
\frac{M_{P}}{\sqrt{1-\frac{R_{s}}{R}}}.
\end{equation}
This energy is related to the uncertainty in determining the value of the position of the shell $R(t)$ by
\begin{equation}
\delta R\cdot\frac{M_{P}}{\sqrt{1-\frac{R_{s}}{R}}}>1,
\end{equation}
showing that at
\begin{equation}
t>2R_{s}\log\left(\frac{R_{0}}{l_{P}}\right)
\label{eq:CollapseTime2}
\end{equation}
the observer will be unable to determine whether the shell has crossed its Schwarzschild radius. The estimate of (\ref{eq:CollapseTime2}) coincides with that of (\ref{eq:CollapseTime}), up to the logarithmic precision.

As we see, once the fluctuations of metric are properly taken into account, the collapse time is no longer infinite (and in fact rather short taking into account how small is the value of the Schwarzschild radius $R_{S}$ for astrophysically interesting scales).

\section{Pre-Hawking Radiation of a Massive Scalar Field}
\label{sec:massive}

In the process of collapsing toward its Schwarzschild radius, the shell emits PHR, losing mass. In principle, if this PHR has sufficient power, the shell could evaporate completely prior to formation of an event horizon, as argued in \cite{VachStojKrauss}. To determine whether collapse can successfully conclude, the collapse time scale as given in (\ref{eq:CollapseTime2}) has to be compared to the time scale of evaporation through PHR emission.  Since the spectrum of the massless PHR is nearly thermal, with the temperature $T$ being of the order of the Hawking temperature  \cite{VachStojKrauss}, the characteristic time scale for the evaporation of the collapsing shell of initial mass $M_0$ through emission of massless quanta coincides with the scale of Hawking evaporation for a black hole with Schwarzschild radius $R_{S}=2GM_0$
\begin{equation}
t_{{\rm evap}}\sim M_{P}^{2}R_{S}^{3}=R_{S}\left(\frac{R_{S}}{l_{P}}\right)^{2}.
\label{eq:masslessevaptime}
\end{equation}
As long as $R_{S}\gg l_{P}$, this time scale is significantly longer than the collapse time given in (\ref{eq:CollapseTime2}), and  PHR cannot prevent the shell from collapse.

Validity of this conclusion strongly depends on the estimate in (\ref{eq:CollapseTime2}) for the collapse time. However, as we shall now argue, even if the estimate in (\ref{eq:CollapseTime2}) is ultimately wrong and an exact solution of the Wheeler-de Witt equation (\ref{eq:WdWfull}) for a collapsing shell reveals a much longer (but finite) collapse time, the conclusion will hold under most (or perhaps all) astrophsically interesting conditions. The ultimate reason for its robustness is the fact that a collapsing shell made of protons, electrons and other constituents of the Standard Model carries global quantum numbers.   These global quantum numbers are not known to be carried by massless particles. Since quanta of  PHR with non-zero global quantum numbers have to be massive,  the rate of their emission is expected to be exponentially suppressed by the usual Boltzmann factor \cite{Hawking}. Unless the Hawking temperature is at or above the mass of the particles, this Boltzmann factor will be a sizable suppression.

In this Section, we shall expand the results of  \cite{VachStojKrauss} to the case of massive radiation, derive its spectrum and analyze its backreaction on the dynamics of a collapsing shell.  In subsequent sections we shall examine the implication of the exponential suppression of massive modes on the likely evaporation pathways of astrophysical black holes.

\subsection{Functional Schr\"odinger Formalism for Massive Pre-Hawking Radiation}
\label{sec:FuncSchrodMassive}

Our goal in this section is to expand the results of \cite{VachStojKrauss} to include the case of massive quanta radiated during the time of collapse. As a result, we will therefore carefully follow their development and notation. 

Let us consider a spherically symmetric configuration of a massive real  scalar field $\Phi$ which propagates in the background of the collapsing shell. The action for the scalar field is
\begin{equation} \label{S}
  S=\int d^4x\sqrt{-g}\frac{1}{2}\left(g^{\mu\nu}\partial_{\mu}\Phi\partial_{\nu}\Phi+m^2\Phi^2\right),
\end{equation}
where $g_{\mu \nu}$ is the background metric given by (\ref{metricexterior}) and (\ref{metricinterior}). The field $\Phi$ can always be expanded into a complete set of eigenmodes
\begin{equation}
  \Phi=\sum_ka_k(t)f_k(r)
  \label{mode expansion}
\end{equation}
such that the Hamiltonian is a simple sum of terms. The total wavefunction then factorizes and can be found by solving a time-dependent Schr\"{o}dinger equation of just one variable.

From (\ref{metricexterior}) and (\ref{metricinterior}), one can split the action (\ref{S}) into interior and exterior parts

\be
  S_{in} \!\!= \!2\pi\!\!\int \!\!dt\!\!\int_0^{R(t)}\!\!\!\!drr^2\dot{T}\Big{[}\frac{-(\partial_{t}\Phi)^2}{\dot{T}^2}+(\partial_r\Phi)^2+m^2\Phi^2\Big{]}
 \ee
and
\begin{align}
  S_{out}\!=\!2\pi\!\!\int \!\!dt\int_{R(\tau)}^{\infty}\!\!drr^2\Big{[}\frac{-(\partial_{t}\Phi)^2}{1-R_s/r}
  &+\left(1-\frac{R_s}{r}\right)(\partial_r\Phi)^2\nonumber\\
  &+m^2\Phi^2\Big{]}\,,
  \label{full action}
\end{align}
where we used (\ref{bigTandlittlet}). It was shown in  \cite{VachStojKrauss} that (\ref{bigTandlittlet}) is given by
\be
  \dot{T}=\frac{dT}{dt}=B\sqrt{1+(1-B)\frac{R^4}{h^2}}
  \label{Tdot}
\ee
where $h=M/(4\pi\mu)$ and $\mu=\sigma(1-2\pi\sigma GR_S)$.

The regime of  most interest is the one when the radius of the shell approaches the Schwarzschild radius $R_s$.

We see from (\ref{Tdot}) that in the near horizon limit, $\dot{T}\sim B\rightarrow0$. Therefore the kinetic term for $S_{in}$ diverges as $(R-R_S)^{-1}$ in this limit. The kinetic term in $S_{out}$ diverges logarithmically, so the $S_{in}$ kinetic term is dominant term. Similarly the potential term in $S_{in}$ vanishes while the potential term in $S_{out}$ becomes finite, so the potential term in $S_{out}$ dominates. Therefore we can write the action as
\begin{align}
          \label{action}
  S \approx 2\pi\int dt\Big{[}&-\frac{1}{B}\int_{0}^{R_s}drr^2(\partial_{t}\Phi)^2\nonumber\\
  &+\int_{R_s}^{\infty}drr^2\left(1-\frac{R_s}{r}\right)(\partial_r\Phi)^2\\
  &+m^2\int_{R_s}^{\infty}drr^2\Phi\Phi\Big{]}\,,\nonumber
\end{align}
where we have changed the limits of integration from $R(t)$ to $R_s$ since this is the region of interest.

Using the expansion in the modes (\ref{mode expansion}), we can rewrite the action as
\begin{align}
  S \approx \int dt\Big{[}&-\frac{1}{2B}\dot{a}_k(t)\mathbf{M}_{kk'}\dot{a}_{k'}(t)+\frac{1}{2}a_k(t)\mathbf{N}_{kk'}a_{k'}(t)\nonumber\\
  &+\frac{m^2}{2}a_k(t)\mathbf{P}_{kk'}a_{k'}(t)\Big{]}
  \label{MatrixAction}
\end{align}
where $\dot{a}=da/dt$, and $\mathbf{M}$, $\mathbf{N}$ and $\mathbf{P}$ are matrices that are independent of $R(t)$ and are given by
\begin{align}
  \mathbf{M}_{kk'}&=4\pi\int_{0}^{R_s}drr^2f_k(r)f_{k'}(r)\label{M}\\
  \mathbf{N}_{kk'}&=4\pi\int_{R_s}^{\infty}drr^2\left(1-\frac{R_s}{r}\right)f_k'(r)f_{k'}'(r)\\
  \mathbf{P}_{kk'}&=4\pi\int_{R_s}^{\infty}drr^2f_k(r)f_{k'}(r).
\end{align}
To simplify (\ref{MatrixAction}), we can write
\be
  S\!=\!\!\int \!\!dt\frac{1}{2}\left[-\frac{1}{B}\dot{a}_k(t)\mathbf{M}_{kk'}\dot{a}_{k'}(t)+a_k(t)\mathbf{R}_{kk'}a_{k'}(t)\right]
  \label{Action}
\ee
where the matrix $\mathbf{R}$ is defined as
\be
  \mathbf{R}_{kk'}=\mathbf{N}_{kk'}+m^2\mathbf{P}_{kk'}
  \label{MatrixR}.
\ee

From the action (\ref{Action}), we can find the Hamiltonian and, according to the standard quantization condition,
\begin{equation}
  \Pi_k=-i\frac{\partial}{\partial a_k(t)}
\end{equation}
is the momentum operator conjugate to $a_k(t)$. From (\ref{M}) and (\ref{MatrixR}), we see that the matrices are Hermitian, therefore it is possible to use the principle axis transformation to simultaneously diagonalize them (see Sec. 6-2 of \cite{Goldstein} for example). Then for a single eigenmode, the Hamiltonian takes the form
\begin{equation}
  H=-\frac{1}{2m}\left(1-\frac{R_s}{R}\right)\frac{\partial^2}{\partial b^2}+\frac{1}{2}Rb^2
  \label{H_b}
\end{equation}
where $m$ and $R$ denote eigenvalues of $\mathbf{M}$ and $\mathbf{R}$, while $b$ is the amplitude of the eigenmode. Hence the Hamiltonian for a single eigenmode takes (\ref{H_b}) the form of a harmonic oscillator with a time-dependent mass.

To find the spectrum of the massive radiation, we must determine the occupation number of the quanta induced during the collapse. From the functional Schr\"odinger formalism, the wave function $\psi(b,t)$ must satisfy
\bd
i\frac{\partial\psi}{\partial t} =H \psi,
\ed
or  from (\ref{H_b})
\be
  \left[-\frac{1}{2m}\left(1-\frac{R_s}{R}\right)\frac{\partial^2}{\partial b^2}+\frac{1}{2}Rb^2\right]=i\frac{\partial\psi}{\partial t}.
  \label{b Schrod}
\ee
Re-writing (\ref{b Schrod}) in the standard form we obtain
\begin{equation}
  \left[-\frac{1}{2m}\frac{\partial^2}{\partial b^2}+\frac{m}{2}\omega^2(\eta)b^2\right]\psi(b,\eta)=i\frac{\partial\psi(b,\eta)}{\partial\eta}
  \label{new b Schrod}
\end{equation}
where
\begin{equation} \label{omega_sq}
  \omega^2(\eta)=\frac{R}{mB}\equiv\frac{\omega_0^2}{B}
\end{equation}
and
\begin{equation}
  \eta=\int dt'B.
  \label{Eta}
\end{equation}
Hence, instead of considering a time-dependent mass we consider the time dependent frequency. In (\ref{omega_sq}) we defined $\omega_0^2 \equiv R/m$. Here we will take  the angular frequency $\omega_0$ to be
\be
  \omega_0=\sqrt{k^2+m^2}\,,
  \label{om_0}
\ee
which is the angular frequency for a relativistic particle. The exact solution to (\ref{new b Schrod}) is given in \cite{Dantas}
\begin{equation}
  \psi(b,\eta)=e^{i\alpha(\eta)}\left(\frac{m}{\pi\rho}\right)^{1/4}\exp\left[\frac{im}{2}\left(\frac{\rho_{\eta}}{\rho}+\frac{i}{\rho^2}\right)b^2\right]
  \label{wavfunc}
\end{equation}
where $\rho_{\eta}=d\rho/d\eta$ and $\rho$ is given by the real solution of the non-linear auxiliary equation
\begin{equation}
  \rho_{\eta\eta}+\omega^2(\eta)\rho=\frac{1}{\rho^3}
  \label{rho eq}
\end{equation}
with initial conditions
\begin{equation}
  \rho(0)=\frac{1}{\sqrt{\omega_0}},\hspace{3mm} \rho_{\eta}(0)=0.
  \label{rho IC}
\end{equation}
The phase $\alpha$ is given by
\begin{equation}
  \alpha(\eta)=-\frac{1}{2}\int_0^{\eta}\frac{d\eta'}{\rho^2(\eta')}.
  \label{alpha}
\end{equation}
Complete information about the radiation in the background of the collapsing shell is contained in the wavefunction (\ref{wavfunc}).

If we consider that an observer at infinity will register quanta of the field $\Phi$ at different frequencies, (\ref{H_b}) tells us that the observer will interpret the wavefunction of a given mode at some later time in terms of simple harmonic oscillator states, $\{\varphi_n\}$, at the final frequency, $\omega_f$. From (\ref{R}) we see that in the near horizon, i.e.~late time limit, $B\sim e^{-t/R_S}$. Therefore we can write the final frequency as
\be
 \omega_f=\omega_0e^{t_f/2R_S}.
 \label{omega_f}
\ee
However, this is in terms of the conformal time, since the derivative in (\ref{new b Schrod}) is in respect to $\eta$ not $t$. (\ref{Eta}) tells us that the frequency in $t$ is $B$ times the frequency in $\eta$, and at time $t_f$, this implies that the observed physical frequency is then
\be
  \omega^{(t)}=B\omega_f\approx e^{-t_f/R_s}\omega_f=\omega_0e^{-t_f/2R_s}
  \label{omega_t}
\ee
where the superscript $(t)$ on $\omega$ refers to the fact that this frequency is with respect to $t$. The initial ($t=0$) vacuum state for each modes is then simply the simple harmonic oscillator ground state
\be
\varphi (b) = \left(\frac{m\omega_0}{\pi} \right)^{1/4} e^{-m\omega_0 b^2/2} \,
\label{varphi_b}
\ee
where $\omega_0=\omega(t=0)$. As mentioned previously, the number of quanta in each mode can be evaluated by decomposing (\ref{wavfunc}) in terms of the simple harmonic oscillator states and computing the corresponding occupation number per mode. The wavefunction for a given mode in terms of simple harmonic oscillator basis  is given by
\begin{equation}
   \psi(b,t)=\sum_nc_n(t)\varphi(b)
\end{equation}
where
\begin{equation}
  c_n=\int db\varphi^{*}(b)\psi(b,t)
  \label{c num}
\end{equation}
which is an overlap of the wavefunction at some later time $\psi(b,t)$ with the simple harmonic oscillator ground state (\ref{varphi_b}). The occupation number at the final frequency $\omega_f$ is then given by
\begin{equation}
  N(t,\omega_f)=\sum_n n |c_n|^2.
\end{equation}
The occupation number in the mode $b$ is then given by
(see Appendix B of \cite{VachStojKrauss})
\begin{equation} \label{N}
  N(t,\omega_f)=\frac{\omega_f\rho^2}{\sqrt{2}}\left[\left(1-\frac{1}{\omega_f\rho^2}\right)^2+
  \left(\frac{\rho_{\eta}}{\omega_f\rho}\right)^2\right] \, .
\end{equation}

\subsection{Numerical Results}
\label{sec:numresults}

In the following plots, we have that the massless case (i.e.~$m=0$) is given by the blue line, $m=5R_s$ is given by the purple line and $m=50R_S$ is given by the brown line.

\begin{figure}[htp]
\includegraphics{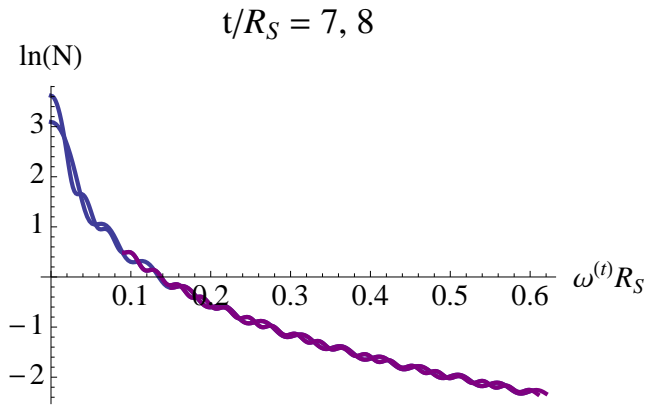}
\caption{$\ln(N)$ versus $\omega^{(t)}R_S$ for $t/R_S=7,8$. Since $\omega^{(t)}=\sqrt{k^2+m^2}e^{-t/2R_S}$, for the massive case $\omega^{(t)}$ has a minimum value which is equal to $me^{-t/2R_S}$, as can be seen here. Note that the occupation for the massless case and the massive case overlap each other, since we are plotting versus $\omega^{(t)}$, which is the same for both cases after the minimum value is obtained. The occupation number grows as $t/R_S$ increases}\label{LnNvswR}
\end{figure}

We have numerically evaluated the spectrum of mode occupation numbers at  finite time and show the results in Figure \ref{LnNvswR} for different values of $t/R_S$. We see first that the occupation number of the massive radiation vanishes below $\omega^{(t)}=me^{-t/2R_S}$, as expected from (\ref{om_0}). Above that, the spectra of the massless and massive cases are identical, since at this point the spectrum depends on $\omega^{(t)}R_S$ only. It is also clear that the occupation number is non-thermal for low frequencies. In particular, as in \cite{VachStojKrauss}, there is no singularity in $N$ at $\omega=0$ at finite time for the massless case.  We also observe oscillations in $N(\omega)$  for both the massless and massive cases.

\begin{figure}[htp]
\includegraphics{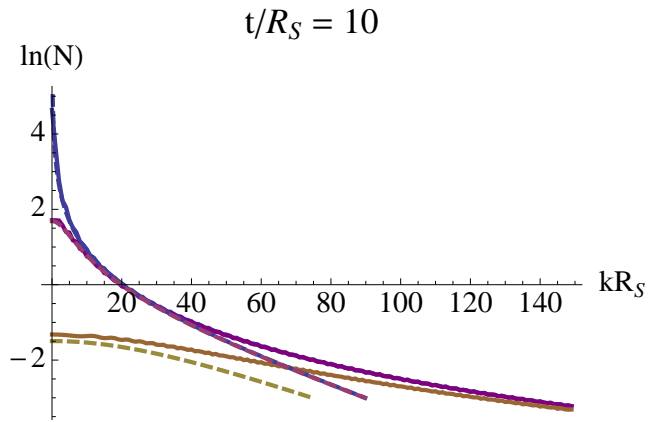}
\caption{$\ln(N)$ versus $kR_S$ for $t/R_S=10$. The dotted lines correspond to the expected Planck distributions (\ref{N_P}). For small values of $kR_S$ the numerical values are similar, however as $kR_S$ increases the results differ. This is due to the non-thermal feature of the radiation for early times.}
\label{LnNvsk10}
\end{figure}

We are interested in the effect that mass has on the spectrum of the pre-Hawking radiation given off during the time of collapse. We therefore plot $\log(N)$ versus $kR_S$ in Figure \ref{LnNvsk10} for different fixed values of $t/R_s$. Here we see that for small values of $kR_S$ the occupation number is greater for smaller masses (largest for the massless case). However, for larger values of $kR_S$ the occupation of each of the masses begin to overlap, as expected from (\ref{omega_t}), since as $kR_S$ increases $\omega^{(t)}$ becomes dominated by $k$ and the effect of the mass is then unimportant. The occupation number of the massive case is obviously suppressed when compared to the massless case. We wish to determine by how much the massive radiation is suppressed.

\begin{figure}[htp]
\includegraphics{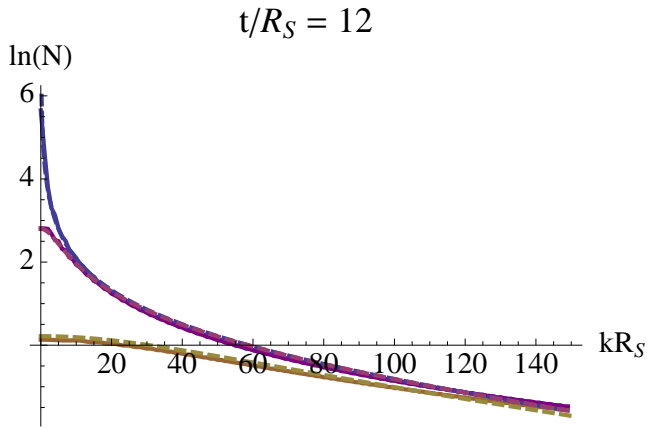}
\caption{$\ln(N)$ versus $kR_S$ for $t/R_S=13$. The dotted lines correspond to the expected Planck distributions. Here the numerical results and expected Planck distributions are in agreement. This is due to the fact that as $t/R_S$ increases the spectrum becomes more and more thermal.}
\label{LnNvsk13}
\end{figure}

To determine the amount of suppression, we compare the spectrum of mode numbers versus $kR_S$ with the occupation numbers for the massless and massive Planck distribution, which are given by
\be
  N_P(\omega)=\frac{1}{e^{\beta\omega}-1}
  \label{N_P}
\ee
where $\omega=|k|$ for the massless case and $\omega=\sqrt{k^2+m^2}$ for the massive case, and where $\beta$ is the inverse temperature. In Figures \ref{LnNvsk10} and \ref{LnNvsk13} we plot $\ln(N)$ versus $kR_S$ for fixed values of $t/R_S$. Before we analyze the results we should note that the value of $\beta$ in Figure \ref{LnNvsk10} is different than that in Figure \ref{LnNvsk13}. This is because here we are plotting versus $kR_s$ not $\omega^{(t)}R_S$, hence we are ignoring the time dependence of the frequency, however keeping the time-dependence of the occupation number. Therefore each moment in time will have a different $\beta$ associated with it.

Figure \ref{LnNvsk10} shows that for small values of $kR_S$ the numerical distributions for the occupation number given in (\ref{N}) are similar to the expected Planck distribution, shown here as dotted lines. However, as $kR_S$ increases, the two distributions differ from each other. This is due to the non-thermal nature of the radiation.

Figure \ref{LnNvsk13} shows that for small values of $kR_S$ the numerical distributions for the occupation number given in (\ref{N}) are in agreement with the expected Planck distribution, again shown here as dotted lines. This is due to the fact that as $t/R_S$ increases, the spectrum becomes more and more thermal for low frequencies, since we are ignoring the red-shift of the frequency.

From (\ref{N_P}) we see that for $k<<m$ the occupation number is exponentially suppressed by $e^{-\beta m}$. Due to the agreement in Figure \ref{LnNvsk13} between the theoretical expected result (\ref{N_P}) and the functional Schr\"odinger equation, we can also conclude that when considering massive radiation, the spectrum is exponentially suppressed by the mass. This result is in agreement with Hawking's prediction (see \cite{Hawking}).

\subsection{Backreaction}
\label{sec:backreaction}

We are finally ready to estimate the rate of pre-Hawking evaporation of the collapsing shell through the emission of quanta with mass $m$. The behavior of the mass $M$ of the shell is given by
\begin{equation}
\frac{dM}{dt}=\frac{1}{2G}\frac{dR_{S}}{dt}=-4\pi R^{2}\left(\frac{m}{8\pi^{2}R_{s}}\right)^{3/2}me^{-4\pi mR_S}.
\label{eq:Rsevap}
\end{equation}
Taking into account (\ref{Rdot}), with solution (\ref{R}) in the near horizon regime (where the pre-Hawking emission is mostly effective), one can estimate 
\begin{equation}
t_{{\rm evap}}\approx\frac{1}{m\sqrt{8\pi mR_{S,0}}}\exp\left(4\pi mR_{S,0}\right)
\label{eq:EvapTimeMassive}
\end{equation}
at $mR_{S,0}\gg1$.
Here $R_{S,0}$ is the Schwarzschild radius at the beginning of the collapse.

As we explained in the Section \ref{sec:quantum}, when the radius of the shell $R$ becomes sufficiently close to $R_S$, so that one can use the expression in (\ref{R}), one naturally expects the perturbation theory in powers of $G$ for the Wheeler-de Witt equation in (\ref{eq:WdWfull}) to break down. This effect is apparently stronger than  PHR, which is entirely contained in the order $G^0$ of the expansion corresponding to the approximation of the functional Schr\"odinger equation (note though that the backreaction of the  PHR is \emph{not} contained in the order $G^0$). This effect is very well known in physics of black holes, where the effective Unruh temperature of the Hawking radiation increases when the observer approaches the event horizon. We shall however assume that the conservative lower bound for evaporation time in (\ref{eq:EvapTimeMassive}) holds in all orders in $G$, since one expects higher orders of $G$ to describe gravitational interaction (capture) of  PHR with the shell as well as gravitational scattering of quanta of  PHR on each other.

Rewriting (\ref{eq:EvapTimeMassive}) in terms of the initial mass of the shell, $M_0$,
\be
t_{{\rm evap}}\sim R_{S}\left(\frac{M_{P}^{2}}{mM_0}\right)^{3/2}\exp\left(\frac{4\pi mM_0}{M_{P}^{2}}\right) \,.
\label{eq:massiveevaptime}
\ee
We see that for $M_0 \gg M_P^2/4\pi m$,
\be
t_{{\rm evap}} \gg R_{S}\left(\frac{M_0}{M_{P}}\right)^{2}\gg R_{S}
\label{eq:massiveevaptime_b}
\ee
The condition $M_0 \gg M_P^2/4\pi m$, can be rewritten to place it more clearly into an astrophysical
context:
\be
M_0  \gg 10^{-11} M_{\rm solar}\frac{\rm eV}{m}
\ee

Note that among different global quantum numbers, baryon number $B$ (and lepton number $L$) is of the most importance. One can naturally expect that the main contribution to the mass of the collapsing shell comes from masses of constituent protons and neutrons. As long as the pre-Hawking emission of particles carrying the baryon number remains suppressed, the shell cannot lose its mass effectively through the pre-Hawking emission of massless quanta. For such shells with masses larger than $10^{11}$ kg, the pre-Hawking evaporation time is given by the time scale in (\ref{eq:massiveevaptime}) rather than that in (\ref{eq:masslessevaptime}). For shells of solar mass $10^{30}$ kg the time scale involved will be of the order of $\exp(10^{19})$ years, apparently much longer than any physically interesting time scale in the problem.

\section{Effect of Baryon and Lepton Number Violation}
\label{sec:BLviolation}

In the previous section, we have shown that the need to emit massive particles in order to radiate away global quantum numbers, such as the baryon or lepton numbers of the shell (or equally, gauge quantum numbers, such as the electro-magnetic charge of the shell), have the potential to enormously extend the evaporation timescale of the shell.   So far, however, we have not included the possibility that physical processes which violate such global quantum numbers could allow a significant fraction of the mass of the shell to be radiated in massless quanta, leaving behind a less massive shell with a higher effective Hawking temperature.

Indeed, the Standard Model contains a mechanism for changing the baryon and lepton numbers of the shell --  non-perturbative electroweak baryon and lepton number violation \cite{Farrar,Sarma,DaiLueStarkStoj}. This process, which can, for example, convert 9 quarks into 3 anti-leptons \cite{Manton}, can allow the baryon number of our shell to relax.  However, the process conserves $B-L$, indeed, it conserves $B/3- L_e$, $B/3- L_\mu$ and $B/3- L_\tau$ separately. (Here $L_i$ is the lepton number of the $i$th generation.) Thus, the minimum mass to which a shell of baryon number B (and lepton number L $\ll$ B) can evaporate by the emission  of massless quanta is not $\sim B m_B$, as one might have guessed, but rather
\be
M_{min} (B) \simeq \frac{B}{3} (m_{\nu_e} + m_{\nu_\mu} + m_{\nu_\tau}) \simeq \frac{B}{3}  \max_i m_{\nu_i} \,.
\ee
In other words
\be
M_{min} \simeq \frac{M_0}{3} \frac{ \max_i m_{\nu_i}}{m_B} \,.
\ee

We should note, that the electroweak non-perturbative baryon and lepton number violation is suppressed by the incredible $e^{8\pi/\alpha}\simeq e^{3000}$ at temperatures below the electroweak symmetry breaking scale. Thus in order for it to affect the evaporation of the shell materially, the electroweak symmetry must be restored.  One might expect that this could happen in the region just above the surface of the shell as it reaches $R_s$. We are going to address this issue in more details in the future work.  


If non-perturbative electroweak baryon number violation does indeed convert the shell's baryons to anti-neutrinos, and thus reduce the mass by many orders of magnitude, neutrino flavor mixing could reduce it still further. Since $\nu_e$, $\nu_\mu$ and $\nu_\tau$ are not mass eigenstates,  it could in fact  be that
\be
M_{min}  \simeq  \frac{M_0}{3} \frac{ \min_i m_{\nu_i}}{m_B} \,.
\ee
where now $i$ enumerates the (three or more) mass eignestates.
In this case we can rewrite
\be
t_{{\rm evap}}\!\sim\! R_{S}\left(\frac{3 m_B M_{P}^{2}}{(m_\nu^{min})^2M_0}\right)^{3/2}\!\!\!\!\!\exp\!\!\left(\frac{4\pi (m_\nu^{min})^2 M_0    }{3 m_B M_{P}^{2}}\right) \,.
\label{eq:massiveevaptimeBLvioln}
\ee
In this case, $t_{{\rm evap}}\gg R_S$ only if
\be
M_0 \gg \frac{3 m_B M_{P}^{2}}{4\pi (m_\nu^{min})^2    } \simeq 10^{7} M_{solar} \left(\frac{eV}{m_{\nu}^{min}}\right)^2.
\ee
Interestingly, with $m_\nu^{min} \simeq 0.1$eV this is likely to mean that the largest observed black holes could decay by employing these mechanisms. However, once again, it is not clear how neutrino oscillations would operate in this environment, and is probably enormously model dependent.

We thus see that there is a potential for physics beyond the perturbative standard model to act on the shell before it collapses.  In this case, what would avert the collapse of the shell is not predominantly pre-Hawking radiation but the radiation of the high energy byproducts of the baryon-number violating interactions.

\section{Conclusions and Discussion}
\label{sec:conclusion}

As was recently demonstrated in \cite{VachStojKrauss}, a spherically symmetric collapsing shell generally loses energy due to pre-Hawking radiation. The spectrum of this radiation is nearly thermal for modes with momentum $k\gg{}R_S^{-1}\sim\frac{M_P^2}{M}$ larger than the inverse Schwarzschild radius of the shell, while the occupation numbers of low momentum modes remain time-dependent until an event horizon is formed. If the time scale of this pre-Hawking evaporation is shorter than the collapse time, we expect this effect to prevent the formation of the event horizon and therefore lead to the resolution of the information loss paradox. The initial pure quantum state of the collapsing shell will evolve into a pure quantum state of the  PHR, which contains the same amount information as the original state.

The question whether such a scenario can be realized in practice is ultimately related to the hierarchy between two time scales in the problem: the collapse time for the shell and its evaporation time through emission of  PHR. As we have argued, while classically the collapse time is infinite, taking quantum fluctuations of the space-time makes it finite and in fact rather small:
\begin{equation}
t_{\rm coll}\approx t_0 + R_S,
\label{eq:colltime}
\end{equation}
where $t_0$ is the time necessary for the shell to collapse to the near horizon regime given in (\ref{R}), irrelevant for the physics discussed, since pre-Hawking evaporation process is ineffective at $t<t_0$.

The collapse time scale in (\ref{eq:colltime}) is to be compared to the total evaporation time for the collapsing shell due to emission of  PHR. If one has a ``realistic'' collapsing shell made of baryons and leptons, its mass is proportional to (approximately) the total number of baryons composing the shell and therefore to the total baryon number $B$ of the shell. To evaporate the mass of the shell means to carry this baryon number away to infinity, and the relevant evaporation time scale should be associated to the process of emitting quanta of \emph{massive} radiation. As we demonstrated, the spectrum of radiation is exponentially suppressed by the mass $m$ of the emitted quanta, meaning that temperature of the black hole must be greater than that of the mass for that species to be radiated away. This would then imply that during most of the collapse of the massive domain wall, the radiation given off would be that of massless particles. For example, if a proton was to be radiated away during the collapse, the process would be suppressed by a factor of
\bd
  e^{-m_e/T_H}\sim e^{-10^{10}}.
\ed
In other words, it is difficult to have the collapsing matter evaporate before the trapped surface is created.

Since massive radiation is only induced once the temperature of the black hole raises above that of the mass of the radiation, the first massive particle that would be radiated would be  a neutrino, which carries lepton number, not baryon number. The lightest baryon, the proton, is at least ten orders of magnitude heavier, and thus its emission is horrendously suppressed.   The shell can therefore lose its baryon number (and its mass) only if baryon number is violated, say by conversion into leptons. Such processes exist within the standard model, but are suppressed by approximately $e^{-3000}$, unless the electroweak symmetry is restored.  
Moreover, such processes would leave a third of the $B-L$ to be carried by the heaviest neutrino, which may be too heavy to experience unsuppressed PHR.  Thus, within the Standard Model, effective radiation of the shell's mass may rely on all of electroweak symmetry restoration, non-perturbative baryon violation, neutrino oscillations, and PHR of neutrinos happening sufficiently rapidly to evade horizon formation. Grand unified theories contain $B$-violating operators, and $B$ (and even $B-L$) could be violated by gravitational operators above the Planck scale and short circuit some of this complication, but again, the rates of such processes in the extremely thin range of radius that is at GUT or Planck temperatures would need to be studied.  Moreover, the radiation of the shell mass would proceed not so much by PHR as by the emission of high energy particles as  a byproduct of baryon number violating interactions at high energy/temperature.

Finally, let us note that the region where the pair production happens is located within vicinity of the shell (where the Schwarzschild metric describing the exterior of the shell is smoothly connected to the Minkowski space-time in the interior of the shell), and while one particle in the pair is escaped to infinity, the other particle from the pair enters the interior of the shell. Apparently, when the pre-Hawking evaporation process becomes effective, the number density of particles in the interior of the shell grows, and eventually the original approximation of Minkowski space-time in (\ref{metricinterior}) inside the shell breaks down. At this point, as one might naturally expect, (a) the space-time inside the shell will be described by the closed FRW metric and will collapse in finite time, and (b) density fluctuations in the gas of particles of  PHR inside the shell will grow and collapse. This scenario explains how exactly the formation of the trapped surface proceeds in the collapse of a spherically symmetric shell.

\section*{Acknowledgments}

The authors would like to thank A. Tolley for discussions. EG and DP were supported by NASA ATP grant to Case Western Reserve University. EG, DP and GDS were supported by a grant from the US DOE to the theory group at CWRU.

\end{document}